\documentstyle[sprocl,psfig]{article}

\def\I{{\cal I}}

\def\P{{\rm P}}
\def\NP{{\rm NP}}
\def\LC{{\rm LC}}

\def\oneloop{{1 \mbox{-} \rm loop}}
\def\twoloop{{2 \mbox{-} \rm loop}}
\def\tree{{\rm tree}}
\def\QCDtree{{\rm QCD\ tree}}
\def\gravtree{{\rm gravity\ tree}}
\def\Split{\mathop{\rm Split}\nolimits}

\def\spa#1.#2{\left\langle#1\,#2\right\rangle}
\def\spb#1.#2{\left[#1\,#2\right]}
\def\lor#1.#2{\left(#1\,#2\right)}
\def\sand#1.#2.#3{%
  \left\langle\smash{#1}{\vphantom1}\right|{#2}%
  \left|\smash{#3}{\vphantom1}\right\rangle}
\def\sandp#1.#2.#3{%
  \left\langle\smash{#1}{\vphantom1}^{-}\right|{#2}%
  \left|\smash{#3}{\vphantom1}^{+}\right\rangle}
\def\sandpp#1.#2.#3{%
  \left\langle\smash{#1}{\vphantom1}^{+}\right|{#2}%
  \left|\smash{#3}{\vphantom1}^{+}\right\rangle}
\def\sandmm#1.#2.#3{%
  \left\langle\smash{#1}{\vphantom1}^{-}\right|{#2}%
  \left|\smash{#3}{\vphantom1}^{-}\right\rangle}
\def\sandpm#1.#2.#3{%
  \left\langle\smash{#1}{\vphantom1}^{+}\right|{#2}%
  \left|\smash{#3}{\vphantom1}^{-}\right\rangle}
\def\sandmp#1.#2.#3{%
  \left\langle\smash{#1}{\vphantom1}^{-}\right|{#2}%
  \left|\smash{#3}{\vphantom1}^{+}\right\rangle}

\def\eqn#1{eq.~(\ref{#1})}
\def\Eqn#1{Eq.~(\ref{#1})}

\def\fig#1{fig.~{\ref{#1}}}

\newskip\humongous \humongous=0pt plus 1000pt minus 1000pt
\def\caja{\mathsurround=0pt}
\def\eqalign#1{\,\vcenter{\openup1\jot \caja
        \ialign{\strut \hfil$\displaystyle{##}$&$
        \displaystyle{{}##}$\hfil\crcr#1\crcr}}\,}
\newif\ifdtup

\newcounter{eqnumber}
\renewcommand{\theeqnumber}{\arabic{eqnumber}}
\def\equn{
\refstepcounter{eqnumber}
\eqno({\rm \theeqnumber})
}

\begin{document}

\noindent
hep-th/9809163 \hfill SLAC-PUB-7897

\hfill  UCLA/98/TEP/35

\hfill  SWAT/98/194

\title{PERTURBATIVE RELATIONSHIPS BETWEEN QCD AND GRAVITY 
AND SOME IMPLICATIONS\footnote%
{Talk presented by Z.B. at Third Workshop on 
Continuous Advances in QCD, Minneapolis, April 16-19, 1998} }

\author{Z. BERN}

\address{Department of Physics, UCLA, \\Los Angeles, CA 90095
\\E-mail: bern@physics.ucla.edu} 

\author{L. DIXON}

\address{Stanford Linear Accelerator Center, Stanford University,\\
Stanford, CA 94309\\E-mail: lance@slac.stanford.edu}

\author{D.C. DUNBAR}

\address{Department of Physics, University of Wales Swansea,\\
Swansea SA2 8PP, UK\\E-mail: d.c.dunbar@swan.ac.uk}

\author{M. PERELSTEIN}

\address{Stanford Linear Accelerator Center, Stanford University,\\
Stanford, CA 94309\\E-mail: maxim@slac.stanford.edu}

\author{J.S. ROZOWSKY}

\address{Institute for Fundamental Theory,  Department of Physics, University
of Florida, \\Gainesville, FL 32611
\\E-mail: rozowsky@phys.ufl.edu} 

\maketitle\abstracts{We discuss nontrivial examples illustrating that 
perturbative gravity is in some sense the `square' of gauge theory.  
This statement can be made precise at tree-level using the Kawai, 
Lewellen and Tye relations between open and closed string tree amplitudes.  
These relations, when combined with modern methods for computing amplitudes, 
allow us to obtain loop-level relations, and thereby new supergravity 
loop amplitudes.  The amplitudes show that $N=8$ supergravity is less 
ultraviolet divergent than previously thought.  As a different
application, we show that the collinear splitting amplitudes of gravity 
are essentially squares of the corresponding ones in QCD.}

\vskip -.9cm
$\null$

\section{Introduction}

Although QCD and general relativity are similar theories in that they both
possess local symmetries and mediate forces, their Lagrangians are
rather different.  In particular, gravity contains an infinite number
of interaction vertices, whereas QCD contains only three- and four-point
vertices.  In this talk we discuss examples demonstrating that
the perturbative $S$-matrices of gravity and QCD are more closely
related than expected based on their Lagrangians.

The existence of relations between gravity and gauge theory amplitudes may
be understood from string theory. At tree level, Kawai, Lewellen and
Tye~\cite{KLT} (KLT) have given precise relations between closed and open
string theory amplitudes. These relations follow (after deforming
integration contours) from the factorization of a closed string integrand
into the product of two open string integrands, one for left-movers and one
for right-movers.  In the infinite string tension limit, where string
theory reduces to field theory, the KLT relations indicate that
$$
\hbox{gravity} \sim \hbox{(gauge theory)} \times \hbox{(gauge theory)} \,.
\equn\label{GravityYMRel}
$$
In this talk we explain how this relationship can be made precise
at loop level. More importantly, we shall discuss its use in 
acquiring nontrivial information about (super) gravity.
The key to exploiting relation (\ref{GravityYMRel}) is to apply modern
methods for computing amplitudes, including improved cutting methods,
helicity and color decompositions.  (For a discussion of these methods 
and for references, see previous reviews~\cite{ManganoReview,Review}.)

As a simple illustration of the notion contained in
\eqn{GravityYMRel}, we show that splitting amplitudes, which describe
the behavior of the gravity $S$-matrix as the momenta of two external
legs become collinear, are given by products of gauge theory splitting
amplitudes.

Another application that we discuss is an investigation of the divergences
in $N=8$ supergravity, based on recycling similar gauge theory
calculations~\cite{SusyEight}. Our interest in $N=8$ supergravity stems
from the fact that it is expected to be the least divergent of all field
theories of gravity.  Furthermore, its high degree of symmetry
considerably simplifies the analytic structure of amplitudes, allowing for
relatively simple computations.  As an important side benefit, it allows
us to test methods for computing multi-loop amplitudes in more
phenomenological theories such as QCD.

The study of divergences in gravity theories has a long
history~\cite{PureGravityInfinity,HoweStelle}.  Because 
Newton's coupling $G_N = \kappa^2/32\pi$ is dimensionful, the presence of 
an ultraviolet divergence indicates that a theory of gravity is not 
fundamental, and that another type of theory, such as string or $M$
theory, may be required.  Except for the explicit calculation of the two-loop
divergence in pure gravity by Goroff and Sagnotti, and later by van de
Ven, analyses of the divergences have generally been based on determining
the form of potential counterterms, subject to power-counting of loop momenta
and symmetry considerations.  However, it is always possible that the
coefficient of a potential counterterm can vanish, especially if the
full symmetry of the theory is not taken into account.

One-loop amplitudes and divergences in $N=8$ supergravity were first
calculated via string theory~\cite{GSB}.  We have computed the
two-loop $N=8$ supergravity amplitude in field theory, by relating its
unitarity cuts to double copies of the cuts of the corresponding $N=4$
super-Yang-Mills amplitude.  In fact, the two-particle cut calculation
can be iterated to generate part of the amplitude at an arbitrary loop
order.  Based on this evidence, we shall argue that $N=8$ supergravity
is less divergent than previously thought.  In particular, the cut
calculations indicate that in $D=4$ the first divergence in four-point
amplitudes occurs at five loops, contrary to previous expectations of
three loops~\cite{HoweStelle}.  Since superspace power-counting only
places {\it bounds} on allowed divergences, there is no real
contradiction.  While it may seem of little importance whether the
divergence starts at five as opposed to three loops, so long as there
is a divergence, the point we wish to stress is that the relation
(\ref{GravityYMRel}) between gauge theories and gravity theories can
be sharpened and exploited to investigate properties of gravity
theories.

\section{Gravity and Yang-Mills at Tree-Level}

\subsection{Lagrangians}

\begin{figure}[t]
\centerline{\psfig{figure=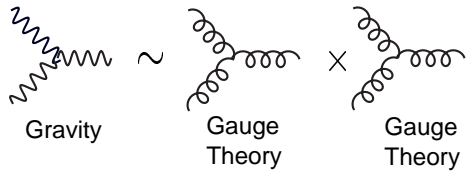,height=1.in}}
\vskip -.4cm 
\caption{String theory suggests that the three-graviton vertex can
be expressed as a product of three-gluon vertices.}
\label{fig:ThreeVertex}
\end{figure}

Before discussing the $S$-matrices, we comment on the Lagrangians of
gravity, ${\cal L}_{\rm gravity} = \sqrt{g} R$, and Yang-Mills, 
${\cal L}_{\rm YM} = - {1\over 4} F^a_{\mu\nu} F^{a\, \mu\nu}$.  
Although the Lagrangians appear to be rather different, \eqn{GravityYMRel}
suggests that the interaction vertices should be related.  In particular, 
one might expect that the gravity three-vertex can be factorized as a 
product of gauge theory three-vertices, as depicted in \fig{fig:ThreeVertex}.
However, such relations do not hold in the standard de Donder (harmonic)
gauge for gravity, in which the three-vertex is~\cite{DeWitt},
$$
G^{\rm harmonic}_{3\mu\alpha,\nu\beta,\rho\gamma}(k_1,k_2,k_3)\ \sim\
k_1\cdot k_2\eta_{\mu\alpha}\eta_{\nu\beta}\eta_{\rho\gamma}\ +\
\hbox{many other terms}\,.
\equn
$$
The exhibited term contains traces over the index pairs of gravitons, 
which prevent the three-graviton vertex from factorizing.

In order for the relation depicted in \fig{fig:ThreeVertex} to hold, one 
has to carefully choose gauges and field variables.  
In particular, in the 
background-field \cite{Background} versions of de Donder gauge for
gravity and of Feynman gauge for QCD, one finds (after color ordering and
stripping the gluon vertex of color factors) that the relation in
\fig{fig:ThreeVertex} does indeed hold~\cite{BDS}.  However, this solution is
not completely satisfactory; it becomes increasingly
obscure to go beyond three points.  Furthermore,
background field gauges are meant for loop effective actions and not for the
(tree-level) $S$-matrix elements.

\begin{figure}[t]
\centerline{\psfig{figure=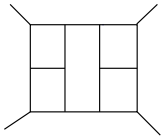,height=.7in}}
\vskip -.4cm 
\caption{An example of a five loop diagram.}
\label{fig:Multiloop}
\end{figure}

In multi-loop gravity Feynman diagram calculations, the number of 
algebraic terms proliferates rapidly beyond the point where computations 
are practical.  Consider the five-loop
diagram in \fig{fig:Multiloop} (which is of interest for ultraviolet 
divergences in $N=8$ supergravity in $D=4$).
In de Donder gauge this diagram contains twelve vertices, each of the
order of a hundred terms, and sixteen graviton propagators, each with
three terms, for a total of roughly $10^{30}$ terms.  Needless to say,
this is well beyond what can be reasonably implemented on any
computer.  Furthermore, standard methods for simplifying diagrams,
such as background-field gauges and superspace, are unfortunately
insufficient for dealing with problems of this complexity.  Direct
string theory based calculations are also not as yet practical
for performing multi-loop calculations, since they are beset with a
variety of technical difficulties.

Our approach will instead be to use cutting methods developed for QCD
computations~\cite{SusyFour,ExactUnitarity,Review} to exploit the
relation~(\ref{GravityYMRel}) and allow us to bypass Feynman
diagram computations.

\subsection{Kawai-Lewellen-Tye Tree-Level String Relations}

At tree level, KLT~\cite{KLT} showed that closed string amplitudes
could be expressed as bilinear sums of open string amplitudes.
The same relations hold for any set of closed string states,
using their Fock space factorization into pairs of open string states.
In the infinite string tension limit, where string theory reduces to 
field theory, $N=8$ supergravity amplitudes are related to $N=4$ 
Yang-Mills amplitudes~\cite{BGK}, making relation~(\ref{GravityYMRel}) 
precise at tree level.  The four- and five-point KLT relations are,
$$
\eqalign{
M_4^\tree (1,2,3,4) & =
     - i s_{12} A_4^\tree (1,2,3,4) \, A_4^\tree(1,2,4,3)\,, \cr
M_5^\tree
(1,2,3,4,5) & =i s_{12} s_{34}  A_5^\tree(1,2,3,4,5)
                                     A_5^\tree(2,1,4,3,5)  \cr
& \hskip 1 cm 
             + is_{13}s_{24} A_5^\tree(1,3,2,4,5) \, 
                           A_5^\tree(3,1,4,2,5) \,,\cr}
\equn\label{KLTExamples}
$$
where $s_{ij} = (k_i + k_j)^2$, the $A_n$ are color-ordered gauge
theory amplitudes, and the $M_n$ are gravity amplitudes.  The arguments
of the amplitudes label the external legs.  For simplicity we have also
suppressed coupling constants and our normalization 
conventions~\cite{SusyEight}.

The tree amplitudes with only external gluons are exactly the same ones
that appear in QCD, because the other fields in the $N=4$ multiplet cannot
appear in intermediate states.  Similarly, the gravity amplitudes are
those of ordinary Einstein gravity.  

Berends, Giele and Kuijf~\cite{BGK} exploited the KLT
relations~(\ref{KLTExamples}) and their $n$-point generalizations to
obtain an infinite set of maximally helicity violating (MHV) gravity tree
amplitudes, using the known MHV Yang-Mills amplitudes~\cite{ParkeTaylor}.
Here we shall explain how one can use the KLT relations to compute
multi-loop gravity amplitudes, starting from gauge theory amplitudes.  
First, though, we discuss a simpler application of the KLT relations: 
the derivation of collinear splitting amplitudes in gravity from those 
in QCD.

\section{Behavior of Gravity Amplitudes for Collinear Momenta.}

QCD helicity amplitudes have a well-known behavior as momenta of external
legs become collinear or soft~\cite{ManganoReview,Review}.  In the case 
of gravity, only the soft limits%
\footnote{The possibility of universal collinear limits for gravity
was noted by Chalmers and Siegel (unpublished).}  
have been discussed in detail~\cite{WeinbergSoftG,BGK}.

At tree-level in QCD, the color-ordered and -stripped amplitudes have
the following behavior as the momenta of legs $1$ and $2$ become
collinear ($k_1 \rightarrow z P$, $k_2 \rightarrow (1-z) P$, and $P =
k_1 + k_2$):
$$
A_n^\tree(1,2,\ldots,n)\ \mathop{\longrightarrow}^{k_1 \parallel k_2}\ 
  \sum_{\lambda = \pm} \Split_{-\lambda}^\QCDtree(1,2) \, 
    A_{n-1}^\tree(P^\lambda,3,\ldots,n)\,,
\equn\label{YMCollinear}
$$
where $\Split_{-\lambda}^\QCDtree(1,2)$ is a splitting amplitude, and
$\lambda$ is the helicity of the intermediate state $P$.  (The other
helicity labels have been suppressed.)
For the pure glue case, one such splitting amplitude is 
$$
\Split_-^\QCDtree(1^+,2^+) = {1\over \sqrt{z (1-z)}} \, {1\over \spa1.2}\,,
\equn\label{YMSplitExample}
$$
where the `$+$' and `$-$' labels refer to the helicity of the gluons, 
$$
\eqalign{
\spa{j}.l &=  \sqrt{2 k_j \cdot k_l}\; e^{i \phi_{jl}}\,, \, \hskip 1.5 cm 
\spb{j}.l  = -\sqrt{2 k_j \cdot k_l}\; e^{-i \phi_{jl}}\,, \cr}
\equn\label{SpinorsDefs}
$$
are spinor inner products, and $\phi_{jl}$ is a momentum-dependent
phase~\cite{ManganoReview}.

From Feynman diagrams (or from the structure of the $n$-point KLT
relations) one can argue that the universal relation~(\ref{YMCollinear}) 
must hold for gravity too~\cite{Future}, 
with $A$ replaced by $M$, and $\Split^\QCDtree$ replaced by a suitable 
gravitational splitting amplitude, $\Split^\gravtree$.  The KLT 
relations~(\ref{KLTExamples}) 
give a simple way to determine $\Split^\gravtree$.  Universality 
permits us to consider any particular collinear limit.  Taking
$k_1\parallel k_2$ in the five-point relation (\ref{KLTExamples}),
we find
$$
\Split^\gravtree(1,2) = 
-s_{12} \times \Split^\QCDtree(1,2) \times \Split^\QCDtree(2,1) \,.
\equn\label{GravTreeColl}
$$
More explicitly, using \eqn{YMSplitExample} for example, we find that
$$
\Split_{-}^{\rm gravity\ tree}(1^+,2^+) =
{ - 1\over z (1-z)} {\spb{1}.{2} \over \spa{1}.{2}}\,.
\equn\label{GravTreeCollExample}
$$
The $s_{12}$ factor has canceled the pole, although a phase singularity 
remains, from the form of the spinor inner products given in
\eqn{SpinorsDefs}; the phase factor $\phi_{12}$ rotates by $2\pi$
as $\vec k_1$ and $\vec k_2$ rotate once around their sum $\vec P$.
The corresponding $4\pi$ rotation in \eqn{GravTreeCollExample} accounts 
for the angular-momentum mismatch of 2$\hbar$ between the graviton $P^+$ 
and the pair of gravitons $(1^+,2^+)$.

In the gauge theory case, the splitting amplitude terms~(\ref{YMCollinear})
dominate the collinear limit; sub-leading behavior is down by a power 
of $\sqrt{s_{12}}$.  In the gravitational analog of \eqn{YMCollinear},
the meaning is different:  There are other terms of the same 
{\it magnitude} as $\spb{1}.{2}/\spa{1}.2$ as $s_{12} \to 0$;
however, these non-universal terms do not acquire any additional 
{\it phase} as $\vec k_1$ and $\vec k_2$ are rotated, and thus they can 
be meaningfully separated from the universal terms.

One application of collinear limits is to help determine the
analytic structure of the graviton $S$-matrix.  In gauge theory, such 
information has been used to find precise expressions for 
$S$-matrix elements~\cite{AllPlus,SusyFour,FourPartons}.  Using the soft 
and collinear properties of gravity we have succeeded in constructing 
Ans\"atze for MHV one-loop amplitudes with an 
arbitrary number of external legs.  These results will be discussed
elsewhere~\cite{Future}, along with a more complete presentation of
the collinear and soft properties of gravity amplitudes.

\section{Multi-Loop Calculations}

Over the years there have been a number of rather impressive
multi-loop Feynman diagram calculations.  However, a number of
important computations remain to be performed.  Two examples of QCD
computations that are required for analyses of experiments, but have
not yet been carried out, are the two-loop contributions to $e^+ \, e^-
\rightarrow 3$ jets and to the Altarelli-Parisi splitting functions.  The
$e^+ \, e^- \rightarrow 3$ jets calculation would be important, for
example, for reducing theoretical errors in the extraction of
$\alpha_s$ from the jet data.  More generally, no computations have
appeared at two and higher loops that involve more than a single
kinematic variable. 

At one loop, a successful recent approach has been to reconstruct
amplitudes from their kinematic poles and cuts~\cite{Review}.  This
approach was used to obtain infinite sequences of one-loop MHV 
amplitudes in QCD \cite{AllPlus} and in supersymmetric
versions of QCD~\cite{SusyFour}, as well as the one-loop helicity
amplitudes for $e^+\, e^- \rightarrow 4$ partons~\cite{FourPartons}.  
Here we will apply the same techniques to two-loop four-point amplitudes
in $N=4$ super-Yang-Mills theory and $N=8$ supergravity.

\subsection{Cutting Methods}

\begin{figure}[t]
\centerline{\psfig{figure=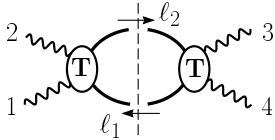,height=.8in}}
\vskip -.4cm 
\caption{The two-particle cut at one loop in the channel carrying 
momentum $k_1+k_2$.}
\label{fig:TwoParticle}
\end{figure}

The cutting method for computing helicity amplitudes has been extensively
discussed for the case of gauge theory amplitudes \cite{SusyFour,Review},
so here we only briefly describe it.  The unitarity cuts of a loop
amplitude are given by phase-space integrals of products of amplitudes
containing fewer loops.  For example, the cut for a one-loop four-point
amplitude in the channel carrying momentum $k_1 + k_2$, as shown in
\fig{fig:TwoParticle}, is given by 
$$
\eqalign{
 \sum_{\rm states} \int  & {d^D \ell_1 \over (4 \pi)^D} \, 
  {i\over \ell_1^2} \,
M_{4}^\tree(-\ell_1,1,2,\ell_2) \,{i \over \ell_2^2}\,
M_{4}^\tree (-\ell_2,3,4,\ell_1) \Bigr|_{\rm cut} \,, 
\cr}
\equn\label{BasicCutEquation}
$$
where $\ell_2 = \ell_1 - k_1 - k_2$, and the sum runs over all states
crossing the cut.  (Polarization labels have been suppressed.)  
We apply the on-shell conditions $\ell_1^2 =
\ell_2^2 = 0$ even though the loop momentum is unrestricted; only
functions with a cut in the given channel are reliably computed in
this way.  (The positive energy conditions are automatically imposed
by the use of Feynman propagators.)

Complete amplitudes are found by combining all cuts into a single
function with the correct cuts in all channels.  If one works with
an arbitrary dimension $D$ in \eqn{BasicCutEquation}, and takes care to 
keep the full analytic behavior as a function of $D$, then
the results will be free of the usual subtraction
ambiguities of cutting methods~\cite{ExactUnitarity,Review}. (The
regularization scheme dependence remains, of course.)

An important advantage of the cutting approach is that the gauge-invariant
amplitudes on either side of the cut can be simplified
{\it before} attempting to evaluate the cut integral~\cite{Review}.
In the case of gravity, we can also make use of the KLT
relations to find convenient representations of the tree amplitudes
for gravity~\cite{SusyEight} in terms of the ones for gauge theory.

\subsection{Maximally Supersymmetric Theories}

The higher degree of symmetry in supersymmetric amplitudes suggests
that they should have a simpler analytic structure than
non-supersymmetric theories such as QCD or Einstein gravity. Therefore,
it is logical to investigate them first.  In particular, 
amplitudes in the maximally supersymmetric theories, $N=4$
super-Yang-Mills and $N=8$ supergravity, should be especially simple;
indeed, this is the case at one loop~\cite{GSB,SusyFour,DN}.
We first discuss multi-loop $N=4$ super-Yang-Mills amplitudes, then 
recycle the answers (using the KLT relations) to get corresponding
results for $N=8$ supergravity, from which we can extract ultraviolet
divergences.

\subsection{$N=4$ Super-Yang-Mills Multi-Loop Amplitudes.}

The key sewing relation used to evaluate the two-particle cuts
for $N=4$ four-point amplitudes is~\cite{SusyFour,BRY,SusyEight},
$$
\eqalign{
\sum_{N=4\ \rm  states}
 A_4^\tree(-\ell_1, & 1, 2, \ell_2) \times
  A_4^\tree(-\ell_2, 3, 4, \ell_1)  \cr
& = - i s t \, A_4^\tree(1, 2, 3, 4) \, 
   {1\over (\ell_1 - k_1)^2 } \, 
   {1\over (\ell_2 - k_3)^2 } \,, 
\cr}
\equn\label{BasicYMCutting}
$$
where all momenta are on shell and the sum is over all states in the
$N=4$ super-multiplet: a gluon, four Weyl fermions and six real scalars. 
The Mandelstam variables are $s= (k_1+k_2)^2$,
$t = (k_1 +k_4)^2$ and $u = (k_1 + k_3)^2$.  \Eqn{BasicYMCutting} may
be easily checked in a helicity basis using four-dimensional
momenta, but it is actually true in all dimensions $(D \le 10)$ and for
all external states belonging to the $N=4$ super-multiplet.

Applying \eqn{BasicYMCutting} to \eqn{BasicCutEquation} at one-loop and 
combining the various cuts immediately yields
$$
A_4^{\oneloop}(1, 2, 3, 4)
 = i \, s t  A_4^{\rm \tree}(1,2,3,4) \, \I_4^{1 \mbox{-} \rm loop}(s,t) \,,
\equn\label{OneLoopYM}
$$
where 
$$
\I_4^{1 \mbox{-} \rm loop}(s,t)  =
\int {d^D \ell \over (2\pi)^D} \, 
{1\over \ell^2 (\ell - k_1)^2 (\ell - k_1 - k_2)^2 (\ell + k_4)^2}\,,
\equn\label{OneLoopIntegral}
$$
in agreement with the previous results of Green, Schwarz and 
Brink~\cite{GSB}.
 
An important feature of the cutting equation (\ref{BasicYMCutting}) is
that the external-state dependence of the right-hand side
is entirely contained in the tree amplitude $A_4^{\rm \tree}$.
This fact allows us to iterate the two-particle cut algebra to {\it all}
loop orders!

\begin{figure}[t]
\centerline{\psfig{figure=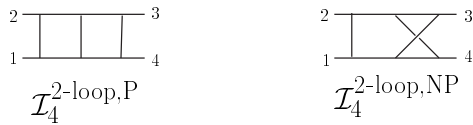,height=.7in}}
\vskip -.4cm 
\caption{The planar and non-planar scalar integrals, 
$\I_4^{\twoloop,\P}(s,t)$ and $ \I_4^{\twoloop,\NP}(s,t)$, appearing in
the two-loop $N=4$ and $N=8$ amplitudes.  Each internal line
represents a scalar propagator.}
\label{fig:PlanarNonPlanar}
\end{figure}

Consider now the two-loop case \cite{BRY}.  The two-loop two-particle
cut sewing algebra is identical to the one-loop case except for the
extra propagators.  The three-particle cuts are more involved, but
generate no other functions beyond those found with two-particle
cuts. After combining all cuts into a single function, a remarkably
simple result emerges for the contribution at leading order in the number 
of colors,
$$
A_{4}^{\LC\ \twoloop}( 1, 2, 3, 4)
= -s t \, A_4^\tree(1,2,3,4) \, \left( s \, \I_4^{\twoloop,\P}(s,t)
+ t \, \I_4^{\twoloop,\P}(t,s) \right) \,.
\equn\label{LeadingColorYM}
$$
The non-planar contributions are also simple~\cite{BRY,SusyEight}. 
The planar and non-planar scalar
two-loop integrals that appear in the amplitudes are shown in
\fig{fig:PlanarNonPlanar}.  Closed-form expressions for the scalar integrals
in terms of known analytic functions are not yet available; nevertheless,
properties such as ultraviolet divergences can be extracted from
\eqn{LeadingColorYM}.

We have compared~\cite{SusyEight} the ultraviolet divergences in 
the above amplitude in $D=7$ and $D=9$ with previous results of
Marcus and Sagnotti~\cite{MarcusSagnotti}.   Up to a minor, unresolved 
discrepancy in the overall normalization of the $D=7$ counterterm, 
we find agreement for both the $D=7$ and $D=9$ counterterms.  
The agreement is rather nontrivial and provides a strong check on our
expressions for the full amplitude.

One may continue to iterate the two-particle cuts to all loop
orders.  We call an integral function that is successively
two-particle reducible into a set of four-point trees `entirely
two-particle constructible'.  Such contributions can be both planar
and non-planar.  For the planar case, i.e. the large $N_c$ 't~Hooft
limit, a simple pattern has been noted~\cite{BRY} that generates the
entirely two-particle constructible contributions.  By extending this
to contributions that require three- or higher particle cuts, one
obtains an ansatz for the form of all large $N_c$ contributions.
(Further details may be found in refs.~\cite{BRY,SusyEight}.)  The 
ansatz has the expected leading-log BFKL~\cite{BFKL} behavior in the 
$s \rightarrow \infty$ limit.  In this limit the gluons dominate, so the
result for $N=4$ super-Yang-Mills agrees with that of QCD.  This
checks that we have the correct ladder diagrams, including
normalizations.

\vskip -.5 cm 
$\null$

\subsection{$N=8$ Supergravity Amplitudes}

Using the KLT four-point relations in \eqn{KLTExamples}, we may
recycle the $N=4$ Yang-Mills sewing equation (\ref{BasicYMCutting})
into an $N=8$ supergravity sewing equation,
$$
\eqalign{
& \sum_{N=8 \rm\ states}  \hskip -.3 cm
M_4^\tree(-\ell_1,  1, 2, \ell_2) \times
  M_4^\tree(-\ell_2, 3, 4, \ell_1) \cr
& =
s^2 \sum_{N=4 \rm\ states}
 \biggl(
  A_4^\tree(-\ell_1,  1, 2, \ell_2) \times
     A_4^\tree(-\ell_2,  3,4 , \ell_1)\biggr)
     \cr
\null & \hskip .8 truecm
\times
\sum_{N=4 \rm\ states} 
\biggl( A_4^\tree(\ell_2, 1, 2 , -\ell_1) \times
                A_4^\tree(\ell_1, 3, 4, -\ell_2) \biggr) \,,\cr}
\equn\label{GravitySewingStart}
$$
where the sum runs over all states in the $N=8$ super-multiplet.
Given the $N=4$ Yang-Mills two-particle sewing 
equation~(\ref{BasicYMCutting}),
it is a simple matter to evaluate \eqn{GravitySewingStart}, yielding
$$
\eqalign{
& \sum_{N=8 \rm\ states}  M_4^\tree(-\ell_1, 1, 2, \ell_2) \times
  M_4^\tree(-\ell_2, 3, 4, \ell_1) \cr \
& \hskip .6 cm
= i stu M_4^\tree(1, 2, 3, 4)
 \biggl[{1\over (\ell_1 - k_1)^2 } + {1\over (\ell_1 - k_2)^2} \biggr] 
\biggl[{1\over (\ell_2 - k_3)^2 } + {1\over (\ell_2 - k_4)^2} \biggr]\,, \cr}
\equn\label{BasicGravCutting} 
$$
where we used
$$
- i \left(s t \, A_4^\tree(1,2,3,4) \right)^2
= s t u  M_4^\tree(1,2,3,4) \,,
\equn\label{OneLoopRelation}
$$
to re-express the prefactor in terms of the $N=8$ tree amplitude.
The sewing equations for the $t$ and $u$ channels are similar to that
of the $s$ channel.

Applying \eqn{BasicGravCutting} at one loop to each of the 
three channels yields the one-loop amplitude,
$$
\eqalign{
{\cal M}_4^{N=8,{\oneloop} }(1, 2, 3, 4)
& =  -i \Bigl( {\kappa \over 2}\Bigr)^4 
s t u  M_4^\tree(1,2,3,4)
 \Bigl(  \I_4^{\oneloop}(s,t)   \cr
& \hskip 2 cm
           + \I_4^{\oneloop}(s,u)  
           + \I_4^{\oneloop}(t,u)  \Bigr) \,, \cr}
\equn\label{OneLoopGravResult}
$$
in agreement with previous results~\cite{GSB}.  We have reinserted the
gravitational coupling $\kappa$ in this expression.  The scalar
integrals are the same ones (\ref{OneLoopIntegral}) appearing in the
$N=4$ Yang-Mills case.

Because the external-state dependence of the right-hand side of 
\eqn{BasicGravCutting} is contained in the tree amplitude,
as in the gauge theory case, the two-loop two-particle cuts are 
given by a simple iteration of the one-loop calculation.
Once again, the three-particle cuts introduce no other functions 
into the amplitude.  Combining the cuts yields the $N=8$ supergravity 
two-loop amplitude~\cite{SusyEight},
$$
\hskip -.3 cm 
\eqalign{ 
 {\cal M}_4^{\twoloop}(1,2,3,4)  & = 
  \Bigl({\kappa \over 2} \Bigr)^6 stu  M_4^\tree(1,2,3,4) 
 \Bigl(s^2 \, \I_4^{\twoloop,\P}(s,t) 
+ s^2 \, \I_4^{\twoloop,\P}(s,u)  \cr
& \hskip  .5 cm  
+ s^2 \, \I_4^{\twoloop,\NP}(s,t)
+ s^2 \, \I_4^{\twoloop,\NP}(s,u) 
 \hskip 0.2  cm 
+\;  \hbox{cyclic} \Bigr) \,, \cr}
\equn\label{TwoLoopGrSquare}
$$
where `$+$~cyclic' instructs one to add the two cyclic permutations of
legs (2,3,4), and $\I_4^{\twoloop,\P/\NP}$ are depicted
in \fig{fig:PlanarNonPlanar}.  These integrals diverge only
for $D\ge 7$; hence the two-loop $N=8$ amplitude is manifestly finite 
in $D=5$ and $6$, contrary to expectations based on superspace 
power-counting arguments~\cite{HoweStelle}.

Since the two-particle cut sewing equation iterates to all loop orders,
one can compute all entirely two-particle constructible contributions, as in
the $N=4$ case.  (The five-loop integral in \fig{fig:Multiloop} falls into
this category.)  Counting powers of loop momenta in these contributions
suggests the simple finiteness formula, $L < 10/(D-2)$, where $L$ is the
number of loops.  This formula indicates that $N=8$ supergravity is finite
in some other cases where the superspace bounds suggest
divergences~\cite{HoweStelle}, e.g. $D=4$, $L=3$.  The first $D=4$
counterterm detected via the two-particle cuts of four-point amplitudes
occurs at five loops, not three loops.  Further evidence that the
finiteness formula is correct stems from the MHV contributions to
$m$-particle cuts, in which the same supersymmetry cancellations occur as
for the two-particle cuts~\cite{SusyEight}.  However, further work is
required to prove that other contributions do not alter the
two-particle-cut power counting.

Another open question is whether we can prove that the five-loop divergence
encountered in the two-particle cuts does not cancel against other 
contributions.  If one could prove that the numerators
of all $N=8$ loop-momentum integrals are squares of the corresponding
ones for $N=4$ Yang-Mills integrals (i.e. they always appear with the
same sign), there would be no need for a detailed investigation of the
cuts.  The iterated two-particle cuts have the required squaring
property, but, as yet, we do not have a more general proof.

\vskip -.5 cm 

$\null$

\section{Conclusions}

\vskip -.1 cm 
Gravity and gauge theories are the two cornerstones of modern
theoretical physics.  In this talk we have discussed nontrivial
examples illustrating that perturbative expansions in gravity theories
are surprisingly similar to those for gauge theories such as QCD,
even though the Lagrangians are rather different.  As one example,
tree-level collinear splitting amplitudes in gravity were shown to be
products of the ones appearing in QCD.  For the case of maximally
supersymmetric theories, where calculations are relatively simple, we
discussed how calculations in $N=4$ super-Yang-Mills can be recycled
to get results for $N=8$ supergravity.  In particular, we obtained the
two-loop four-point amplitudes for each theory 
in terms of scalar integrals.  Furthermore, the two-particle cut 
calculus iterates to {\it all} loop orders.  From these considerations,
it appears that $N=8$ supergravity is less divergent than previously thought.

It would be nice to find a field theoretic reformulation of gravity 
where the connection (\ref{GravityYMRel}) to gauge theory is explicit.
On the more practical side, we are optimistic that the same cutting 
techniques discussed here can be applied to multi-loop amplitudes in 
theories with less supersymmetry, such as the two-loop corrections to 
$e^+ e^- \rightarrow$ 3 jets in QCD.  

\vskip .2 cm 
This research was supported by the US Department of Energy
under grants DE-FG03-91ER40662, 
DE-AC03-76SF00515 and DE-FG02-97ER41029.

\section*{References}

\vskip -.2 cm 

\end{document}